# In-situ electro-optic sampling of microwave signals under cryogenic conditions and for superconducting applications

Shekhar Priyadarshi[1], Rinis Ferizaj[1], Oliver Kieler[1], Alexander Fernandez Scarioni[1], Judith Felgner[1], Abdulrahman Widaa[1], Johannes Kohlmann[1], Thomas Fordell[2], Jaani Nissilä[2], Antti Kemppinen[2], and Mark Bieler[1*]

[1] Physikalisch-Technische Bundesanstalt, Bundesallee 100, 38116 Braunschweig, Germany.

[2] VTT Technical Research Centre of Finland Ltd, 02150 Espoo, Finland.

[*] *Corresponding author: mark.bieler@ptb.de.*

We demonstrate a cryogenic electro-optic sampling (EOS) setup that allows for the measurement of microwave signals at arbitrary positions on a cryogenic chip-scale device. We use a Josephson Arbitrary Waveform Synthesizer (JAWS) to generate quantum-accurate voltage signals and measure them with the EOS setup, allowing for the calibration of its response, yielding traceability of the microwave measurements to a quantum standard. We use the EOS setup to determine the time-domain response of ultrafast cryogenic photodiodes and the electrical reflection coefficient, i.e., the $S_{11}$ scattering parameter, in a superconducting transmission line. Finally, we introduce an optical femtosecond pulse source which can be used to study the fidelity of superconducting transmission lines and terminations, as well as reflections from elements like Josephson junction arrays embedded in them.

**Keywords -** Ultrashort voltage pulse, electro-optic sampling, photodiode characterization, Josephson Arbitrary Waveform Synthesizer (JAWS), cryogenic temperature, fiber-chip coupling, pulsed laser.

## I. Introduction

Most solid-state approaches for quantum computing and many other quantum technologies rely on microwave signals [1,2]. Usually, there is only indirect information about the microwave signal levels and purity at the devices in the cryostat. One of the problems is a tradeoff between thermal insulation and bandwidth of metallic cables that carry microwave signals from the cryogenic devices to room temperature measurement electronics. Hence such cables have significant losses already around 1 GHz. In particular, none of the conventional methods allow reliable time-domain, oscilloscope-like measurements of cryogenic signals in the interesting frequency range between 1 GHz and 1 THz, whereas electro-optic sampling makes those feasible [3–5].

Unlike electrical cables, optical fibers have simultaneously wide bandwidth and low thermal conductivity, which has raised the interest, e.g., in the opto-electrical control and readout of quantum computers and other quantum technologies [6,7]. One of the promising approaches for cryogenic signal generation is the Josephson Arbitrary Waveform Synthesizer (JAWS) that can also be driven optically, and its variant, the Josephson Pulse Generator (JPG), that already enabled qubit control [8–10].

Recently, we demonstrated electro-optic sampling (EOS) of a cryogenic photodiode [11,12], where optical-fiber guided femtosecond laser pulses offer a very high bandwidth, which theoretically exceeds 1 THz, and allow for low-invasive and in-situ measurements. In this context, invasiveness refers to unwanted electrical coupling between the EOS system and the device under test. One interesting application for a broadband EOS setup is to study electric pulse propagation in superconducting circuits, such as JAWS, being sensitive to signal reflections [13,14]. Here we report on a reversed experiment, where JAWS is used as a quantized reference signal source that allows the calibration of the EOS sampling setup. Moreover, we present several application examples in which we use EOS for in-situ measurements of electrical signals and electrical reflection coefficients in superconducting circuits.

The remainder of this paper is structured as follows. In Sec. II we introduce the EOS setup based on femtosecond lasers. In Sec. III we utilize the setup for the characterization of commercially available photodiodes with nominal bandwidths of 20 GHz and 60 GHz, compare their responses at room temperature and at cryogenic temperature, and report on optimization of their time-domain responses with respect to optical excitation power. We extend the EOS setup in Sec. IV to measurements of electrical reflection coefficients, corresponding to $S_{11}$ scattering parameter measurements. We then use a JAWS to calibrate the EOS setup to a quantum standard in Sec. V and measure the JAWS output at cryogenic temperatures in the time-domain in Sec. VI. Finally, we introduce the EOS of an optical pulse pattern generator in Sec. VII before we conclude and give an outlook in Sec. VIII.

## II. Experimental set-up

The main part of the experimental setup is shown in Fig. 1. Different types of commercially available p-i-n photodiodes (PD) are flip-chip bonded to one end of a coplanar waveguide (CPW), which is made of either Nb or Au. The dimensions of the CPW were as follows: 200 µm width of the ground conductors, 29 µm width of the signal conductor, and 18 µm gap between signal and ground conductors. Different CPW lengths have been employed, see further below in this section. For the optical excitation of the PDs we have two pulsed light sources. In the basic characterization measurements of the PD we use pump laser pulses with a duration of 200 fs full width at half maximum (FWHM) and a center wavelength of 1340 nm. For more complex pulse pattern measurements we have developed an optical generator described in more detail in Sec. VII. The generated voltage pulses propagate along the CPW on which we placed a $LiTaO_3$ crystal. The electric-field-induced refractive index change in the $LiTaO_3$ crystal is read out using probe laser pulses with a duration of 200 fs and a center wavelength of 1550 nm [15]. In this regard we like to note that the corresponding phase changes to the probe beam are so small ($\sim 5 \times 10^{-4}$ for a 1 V signal to be measured) that any nonlinearity of the EOS technique can be neglected. Time resolution for these EOS measurements is obtained by changing the temporal delay between the pump and probe pulses. Both pulses are delivered to the chip using single mode optical fibers (SMF) ending in borosilicate ferrules. More information on the experimental setup and the fiber-chip coupling is given in Ref. [11].

The EOS is performed in two different ways. In a first realization, used for the experiments discussed in Sec. III, the $LiTaO_3$ crystal (dimensions 10 mm x 5 mm x 0.05 mm) is glued together with the probe fiber ferrule to the sampling position on the CPW made of Au with an overall length of the CPW of 8.5 mm. The dimensions of the CPW sections as depicted in Fig. 1(b) were, from the left-hand side to the right-hand side, 1.5 mm ($CPW_0$), 5 mm ($CPW_{EO}$), and 2 mm ($CPW_0$). Moreover, the probe beam was incident on the $LiTaO_3$ crystal at 2 mm (3 mm) distance from its left-hand-side (right-hand-side) end.

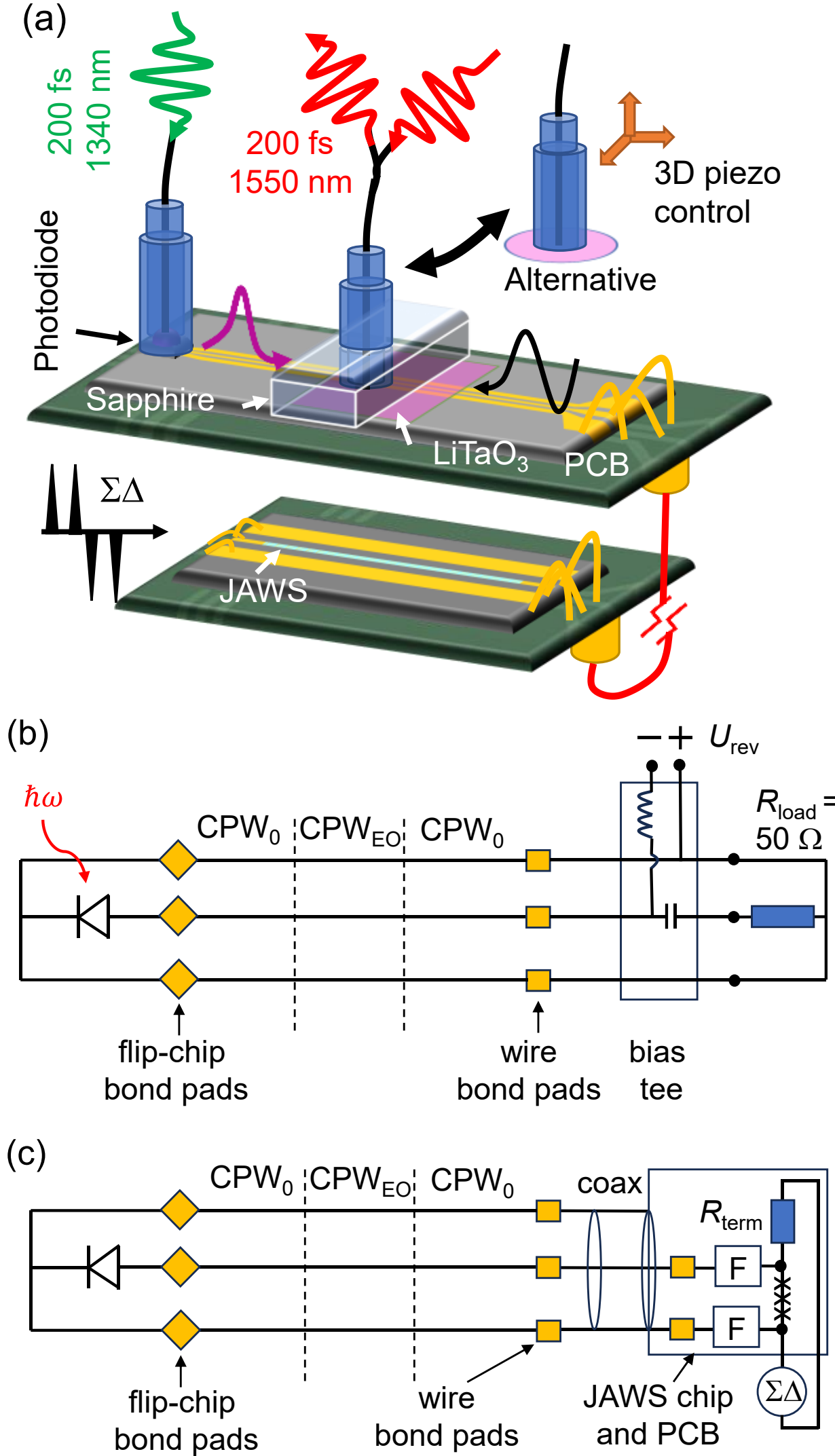

Fig. 1. (a) Overview of the experimental scheme depicting in-situ EOS measurements of voltage pulses and JAWS synthesized signals at a cryogenic temperature. For the EOS measurements without JAWS chip, the right-hand-side of the CPW of the upper chip is terminated with a bias tee outside the chip. For visualization purposes we have simplified the output of JAWS chip, which includes additional low-pass filters across the array that are not shown in (a), also compare with (c). The red line connecting the JAWS chip with the EOS platform denotes a coaxial cable. (b) Schematic block diagram for the EOS experiments using photodiodes. (c) Schematic block diagram for the EOS experiments using JAWS chips. $CPW_0$: coplanar waveguide with nominal $50\,\Omega$ impedance. $CPW_{EO}$: coplanar waveguide region with attached EO crystal on top. The JAWS chip comprises an array of Josephson junctions (JJs) and low pass filters (F) at its output. The JAWS is driven with a ΣΔ-modulated electrical pulse sequence generated by a room-temperature pulse pattern generator and delivered via a coaxial cable to the JAWS circuit.

In a second realization, used for the experiments discussed in Secs. IV to VII, the $LiTaO_3$ crystal (approximate dimensions 2 mm x 2 mm x 0.05 mm) is only glued to the probe fiber ferrule and this arrangement can be moved using a 3D cryogenic piezo stage. The latter allows us to perform spatially resolved EOS measurements. In these measurements the overall length of the CPW made of Nb was 12.5 mm. The dimensions of the CPW sections as depicted in Figs. 1(b) and (c) were, from the left-hand side to the right-hand side, 6 mm ($CPW_0$), 2 mm ($CPW_{EO}$), and 4.5 mm ($CPW_0$). Moreover, the probe beam was incident on the center of the $LiTaO_3$ crystal with approximately 1 mm distance to both ends.

The LiTaO3 crystals have an anti-reflection coating on the top surface and a high-reflectivity coating on the bottom surface. The 1550 nm laser pulses are emitted from the end facet of the fiber, travel through the $LiTaO_3$ crystal, and are back-reflected into the fiber again. From power measurements, we estimate that this process has an approximate efficiency of 40%, i.e., 40% of the incident power is again coupled into the fiber and guided towards the ellipsometric detection/analysis. In the experiments, the incident cw power was approximately 7.5 mW and the back reflected power approximately 3 mW. Due to the high reflectivity coating of the $LiTaO_3$ crystal, we expect that hardly any laser light reaches the Nb or Au CPW or the corresponding Si substrate. Instead, we expect that the reflected laser light that is not coupled into the fiber again, will be absorbed or diffusely reflected by the Cryoperm shielding of the sample. Consequently, we do not expect any significant influence of quasiparticle generation on our measurements. We like to emphasize that EOS could also be performed with significantly less probe power. However, we have not performed a consistent study, yet, in which we analyze the obtained signal-to-noise-ratio versus probe power.

In voltage pulse measurements presented in Secs. III and IV, the other end of the CPW is used to apply the DC bias to the PD via a bias tee, see Fig. 1(b). For the measurements presented in Secs. V the bias to the end of the CPW is replaced with the output of a Josephson Arbitrary Waveform Synthesizer (JAWS) [16,17] and optical excitation of the PD is ceased, see Fig. 1(c). This allows us to measure the JAWS output propagating along the CPW exactly in the same condition employed for the measurement of voltage pulses generated from the PD. Such measurements enable calibration of the voltage pulses to a quantum standard. We use the same experimental geometry in section VI for EO measurement of temporal shape of two different JAWS output signals. The whole arrangement as depicted in Fig. 1 is inserted into a liquid He Dewar such that all measurements are done at 4 K unless otherwise noted.

## III. Characterization of fast photodiodes

As a first application example, we have used the EOS setup for the characterization of the time-domain response of PDs with a nominal bandwidth of 20 GHz and 60 GHz. Being important for the remainder of this manuscript, we present some of the results published already in Ref. [11], in which a more comprehensive study of the PD responses over a large parameter range has been performed. Yet, we also present unpublished results with respect to bandwidth estimation and optimization of the tail-to-peak ratio of the time-domain PD responses.

Figures 2(a) and (b) show the normalized time traces of voltage pulses generated from the 20 GHz and 60 GHz PDs, respectively, and propagating on a CPW made from Au. Sampling was performed at temperatures of 4 K (solid blue) and 300 K (dashed red). During the measurements the reverse bias $U_{rev}$ and the average photocurrent $I_{ph}$ were kept at 2 V (1.2 V) and 20 µA (5 µA), respectively, for the 20 GHz (60 GHz) PD. We observe that both the temporal width and the tail of the voltage pulses decrease for both PDs when reducing the temperature from 300 K to 4 K. Moreover, the temporal width and the tail are reduced for the 60 GHz PD as compared to the 20 GHz PD. The reduced width of the voltage pulses and, in particular, the reduced tail correspond to an improved performance of the PD's time-domain response at cryogenic temperatures.

The FWHM of the measured voltage pulse of the 20 GHz PD at 4 K shown in Fig. 2(a) is 18.8 ps and decreases to 14.2 ps upon reduction of the photocurrent $I_{ph}$ to 5 µA. Using the well-known relation between the FWHM ($t_{\mathrm{FWHM}}$) and 3-dB bandwidth ($f_{3\mathrm{dB}}$) of $t_{\mathrm{FWHM}} \times f_{3\mathrm{dB}} = 0.44$ for Gaussian pulses we obtain bandwidth estimates of ~23 GHz (for $I_{ph}$ = 20 µA) and ~31 GHz (for $I_{ph}$ = 5 µA) for the PD with a nominal bandwidth of 20 GHz. For the PD with a nominal bandwidth of 60 GHz we extract a FWHM at 4 K of 10.4 ps (for $I_{ph}$ = 5 µA) and 5.4 ps (for $I_{ph}$ = 2.3 µA), yielding bandwidth estimates of ~40 GHz and ~80 GHz, respectively. We note that the bandwidths obtained from an FFT of the time-domain measurements shown in Fig. 2 are considerably smaller due to the long tails and strongly depend on the time window used for the Fourier transformation. They can be as low as a few GHz, see Ref. [11]. Therefore, we believe that the analysis based on the time-domain measurements is more appropriate.

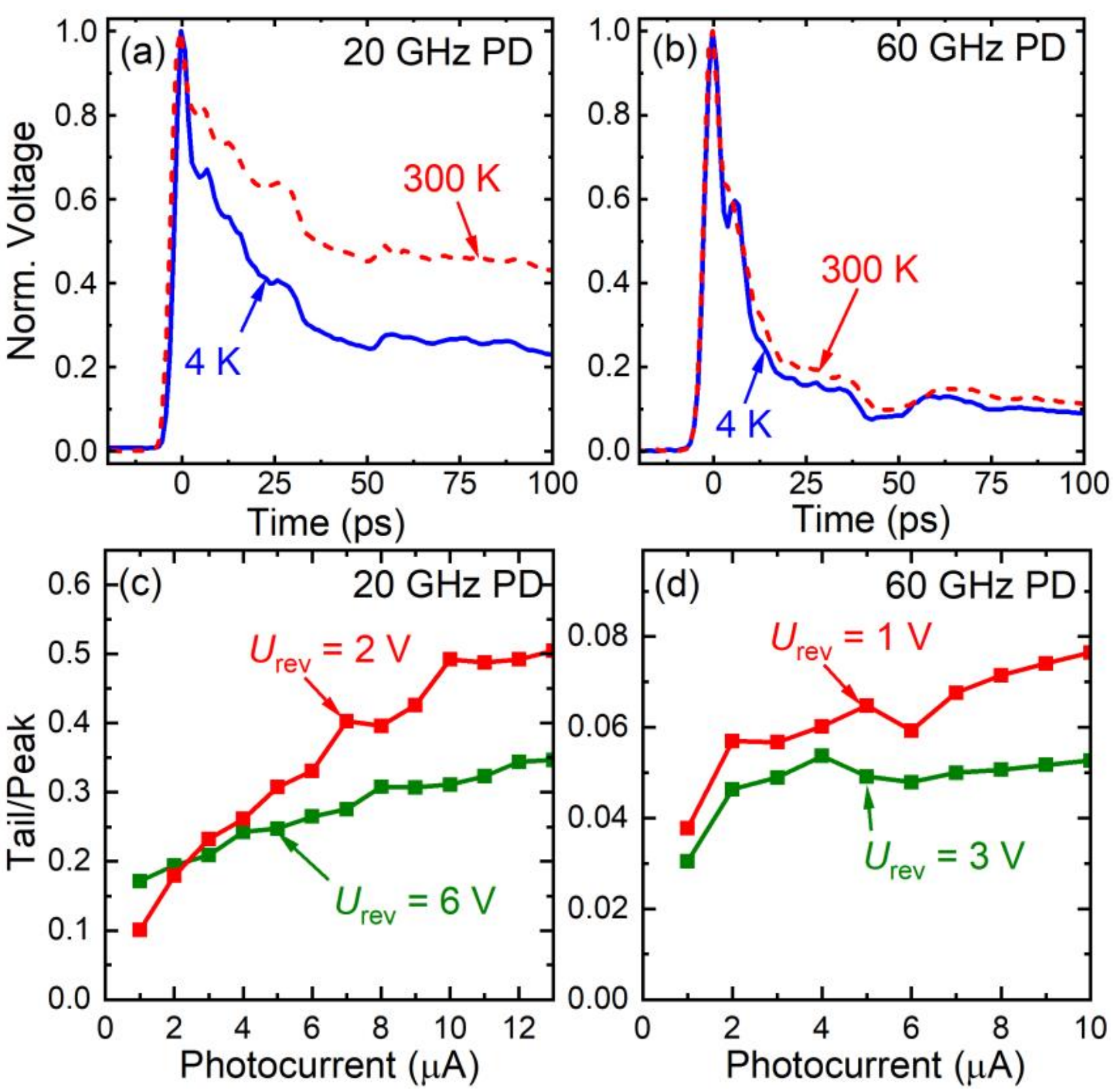

Fig. 2. Temporal responses of a PD with a nominal bandwidth of (a) 20 GHz ($U_{\mathrm{rev}}$ = 2 V, $I_{\mathrm{ph}}$ = 20 μA) and (b) 60 GHz ($U_{\mathrm{rev}}$ = 1.2 V, $I_{\mathrm{ph}}$ = 5 μA), measured by EOS on a AuCPW [10] at temperatures of 4 K and 300 K. (c) and (d) show the dependence of the tail-to-peak ratio of the 20 GHz and 60 GHz PD, respectively, versus photocurrent $I_{\mathrm{ph}}$ and for two different reverse biases $U_{\mathrm{rev}}$. The measurements shown in (c) and (d) were done at 300 K.

As already discussed, the tails of the PD responses are quite significant, especially for the 20 GHz PD. We have therefore studied to which extent the tail can be reduced. Figures 2(c) and (d) show the tail-to-peak ratio of the time-domain responses of the 20 GHz and 60 GHz PD, respectively, at room temperature. Here, the value for the tail has been determined by averaging over the time-domain response between 80 ps and 150 ps after the peak. These plots clearly prove that the tail-to-peak ratio can be reduced by increasing the reverse bias and by reducing the excitation power, the latter of which goes along with a reduction of photocurrent. Here, the 60 GHz PD performs best and the tail can be reduced to a few percent of the peak value over a rather large range of photocurrent. This is beneficial for applications in superconducting technology where short and return-to-zero pulses are required.

The long tail in the time-domain measurements might be the result of femtosecond excitation of the photodiodes. Most likely, the femtosecond excitation leads to electric field screening in the depletion region of the p-i-n-photodiodes, which may significantly decrease the carrier drift velocity, slowing down the temporal response and leading to long tails [18,19]. Another possible reason leading to broadened voltage pulses is absorption saturation in the depletion region. In such a case absorption takes place also in the diffusion region, slowing down the temporal response. In Sec. VII and Fig. 7 we will present additional experimental data strongly supporting these assumptions, as we see that broader optical pulses with less peak energy produce shorter electrical pulses.

## IV. Reflection coefficient in superconducting circuits

When performing EOS measurements over a longer time epoch it will be unavoidable that reflections from transmission line discontinuities or impedance mismatches influence the measurements. Here it should be emphasized that even CPW tapering, bond wires, or coaxial connectors lead to reflections which overlap with the incident signal. In such cases it will be essential to determine the electrical reflection coefficient at the measurement position in order to distinguish between forward and backward propagating signals.

This electrical reflection coefficient can be extracted from EOS measurements at different spatial positions on the CPW: The electrical reflection coefficient corresponds to the $S_{11}$ scattering parameter referred to the impedance of the CPW. In our measurements we employ a CPW made from Nb with an impedance of 50 Ω. Here, we like to note that in the superconducting state and at very low frequencies the impedance of the transmission line stays at 50 Ω and does not become complex as compared to resistive Au transmission lines around and below 1 GHz [20].

The technique of extracting electrical reflection coefficients from two-positions measurements on a CPW has previously been demonstrated at room temperature [20], but, so far, not applied to cryogenic environments. In the following we just briefly sketch the measurement principle. More information can be found in Ref. [20]. We start with the measurement of two voltage pulses, $V_1$ and $V_2$, at two positions on the CPW. The two measurement positions need to be at least 1 mm or 2 mm away from a transmission line discontinuity in forward direction, i.e., in the propagation direction of the voltage pulses. We then cut the measured time traces after a certain time by setting the data points at later times to zero, such that no reflection is present in the time traces. (On a typical CPW, signals propagate with a speed of approximately 1 mm/10 ps. Thus, if a transmission line discontinuity is 1 mm away from the measurement positions, the first reflection will enter the measured time window after 20 ps.). With the cut-off time traces, $V_{1,\mathrm{cutoff}}$ and $V_{2,\mathrm{cutoff}}$, we calculate the transfer function of the CPW between the two measurement positions in the frequency domain $p = V_{2,\mathrm{cutoff}}/V_{1,\mathrm{cutoff}}$. Once, $p$ is known, the reflection coefficient $\Gamma$ at the measurement position two is obtained from

$$\Gamma = \frac{V_2 - pV_1}{p(V_1 - pV_2)}\,, \tag{1}$$

which again denotes an equation in frequency domain.

The aforementioned scheme has been applied to the cryogenic EOS measurements in which we detect voltage pulses produced from a PD. The green and blue trace of Fig. 3(a) show the measured voltage pulses $V_1$ and $V_2$, respectively, with the measurement positions being separated by approximately 600 µm. In the inset of Fig. 3(b) we plot the time-domain reflection coefficient at the second measurement position. This trace corresponds to the inverse Fourier transformation of $\Gamma$ defined by Eq. (1). This reflection shows a large bipolar signal around 120 ps to 140 ps. We identify the end of the Nb CPW and the bond pads/wires as the source for this reflection signal.

With the reflection coefficient it is possible to separate forward and backward propagating signals. The blue and red lines in Fig. 3(b) show the EOS measurement at the second position, $V_2$, and the forward propagating signal at this position, respectively. The signals at ~280 ps and ~530 ps in the blue curve correspond to reflections from the right-hand-side end of the CPW, see Fig. 1, propagating backwards and, thus, are removed in the red curve. In contrast, the signals at ~420 ps and ~670 ps denote reflections from the left-hand side end of the CPW where the PD is flip-chip bonded, see Fig. 1. Since these reflections belong to the forward propagating voltage pulse, they are still visible in the red curve of Fig. 3(b).

## V. Calibration of EOS measurements using JAWS

In general, EOS measurements done in a pump-probe configuration yield voltage signals in arbitrary units (e.g., units of the lock-in amplifier). In such measurements the pump beam (or the signal to be measured) is amplitude-modulated and the pump-beam-induced change of the probe beam is detected with a lock-in amplifier at the modulation frequency of the pump beam. To convert these arbitrary units of the lock-in amplifier to the unit volt, one typically applies a reference signal $V_{\mathrm{ref}}$ with a well-known amplitude to the EOS setup [21]. The EOS response to this reference signal $V_{\mathrm{ref,EOS}}$ then allows to convert other EOS measurements to the unit volt by a simple multiplication with $V_{\mathrm{ref}}/V_{\mathrm{ref,EOS}}$.

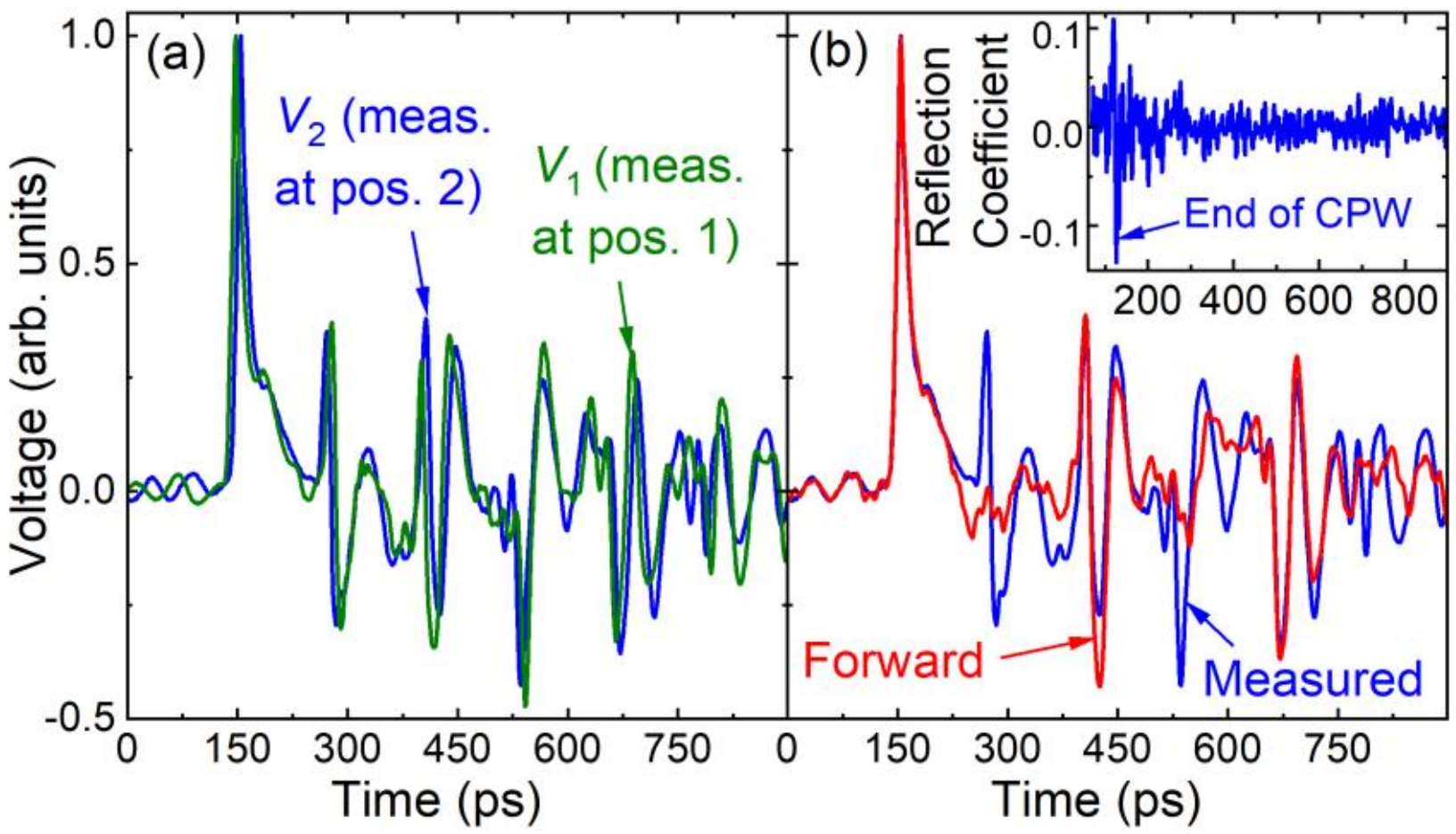

Fig. 3. (a) EOS measurement of two voltage pulses $V_1$ and $V_2$ on a Nb CPW produced from a PD with a nominal bandwidth of 20 GHz. The two measurement positions were approximately 600 µm apart. (b) EOS measurement of $V_2$ (same trace as in (a)) and forward propagating signal at this position. The inset shows the reflection coefficient at the second measurement position. The trace corresponds to the inverse Fourier transformation of Γ defined by Eq. (1).

The response function of EOS is flat at low frequencies and rolls off at very high frequencies >> 10 GHz, see Ref. [20], in which the EOS response function has been experimentally determined for a sampling geometry being similar to the one used in the current manuscript. Thus, it is possible to perform the calibration using a reference signal $V_{\mathrm{ref}}$ at low frequencies (~1 MHz) and this calibration will also be valid for GHz frequencies before the roll-off takes place. When knowing the EOS response function and deconvolving the measured time traces with this function, the calibration will be valid even for all measured frequencies.

In a previous study [21] we have shown that due to the frequency dependence of the optoelectronic devices employed for EOS, the reference signal should be equal to the modulation frequency of the optical pump beam (or of the signal to be measured). Since we modulate the pump beam in our experiment with an acousto-optic modulator at 1 MHz, we also employ a reference signal at this frequency.

To obtain direct traceability of the EOS measurements to a quantum standard, we have used a JAWS for the generation of the reference signal. Our JAWS chip comprises 3,000 Josephson junctions with a characteristic voltage $V_c$ = 17.3 µV and a critical current density $J_c$ = 93 µA/µm². The JAWS chip was driven by a pulse sequence to generate a single- frequency output signal at 1 MHz with a peak-to-peak value of 17.36 mV, see Fig. 4(a). The operation margins (quantum locking range) are approximately 0.4 mA for an output frequency of 1 MHz (and increase to 3 mA for a reduction of the output frequency to 1.75 kHz). As shown in Fig. 1(c), for the EOS experiments, the JAWS chip was directly connected to the right-hand-side end of the CPW. The EOS response ($V_{ref,EOS}$) to the JAWS signal ($V_{ref}$) was measured with the lock-in amplifier. This allowed us to calibrate voltage pulse measurements as presented in Sec. III to the unit volt. Figure 4(b) shows such a calibrated voltage pulse produced from a PD with a nominal bandwidth of 20 GHz and to the best of our knowledge this is the first EOS measurement with direct traceability of the voltage axis to a quantum standard.

The JAWS amplitude error is in the nV range (estimated from previous low-frequency measurements [22,23]). At 1 MHz, deviations ranging between $10^{-2}$ and $10^{-3}$ were reported for AC

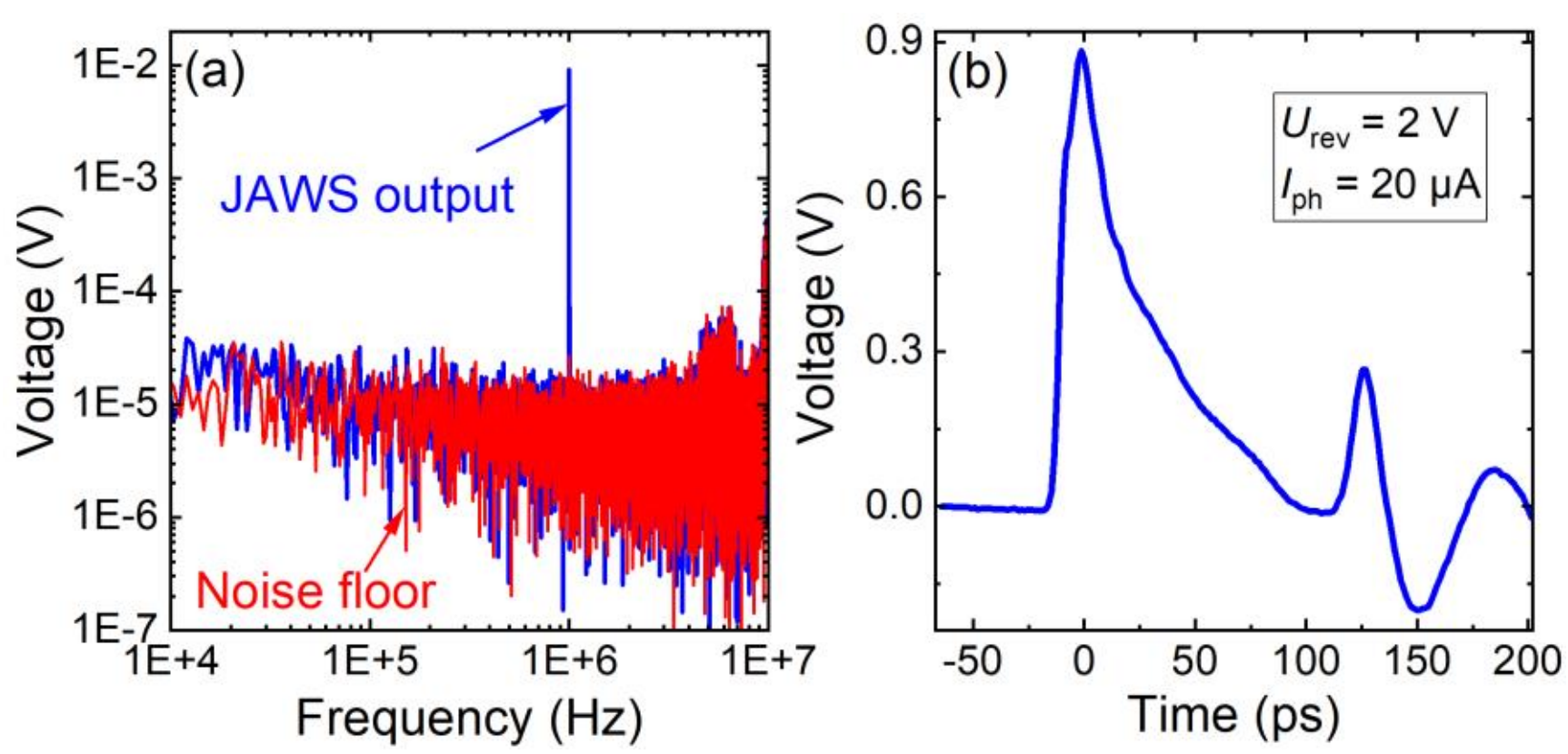

Fig. 4. (a) Spectral amplitude of the JAWS output with a single tone at 1 MHz measured with a conventional spectrum analyzer at room temperature. (b) EOS measurement of a voltage pulse on a Nb CPW produced from a strongly excited PD with a nominal bandwidth of 20 GHz. The y-axis is calibrated to the unit volt using the JAWS output. The oscillation around 150 ps denotes a reflection from the bond pad/wires, see also Fig. 3 and corresponding discussion in the text.

Josephson standards with rather long coaxial cables of 1.5 m [24] and approximately 1 m [25], respectively. With this information, JAWS amplitude errors and signal propagation errors over short coaxial lines can be neglected as compared to laser noise and temperature variations influencing the EOS signal. In our experiments, temperature variations ($\pm$2 K) of room-temperature optics (laser, fibers, other optical components) led to a relative variation < 0.05 of the detected reference signal used for amplitude calibration and this is the current uncertainty of our calibration method.

## VI. Direct EOS of the JAWS output

In the measurements of the reference signal as discussed in the previous section, we obtained one value for $V_{\mathrm{ref,EOS}}$, expressing the peak-to-peak value of the JAWS signal. Of course, another point of interest is the direct time-domain measurement of the JAWS signal. For this experiment we used another JAWS chip with 14,000 Josephson junctions with a characteristic voltage $V_{\mathrm{c}}$ = 17.3 μV and a critical current density $J_{\mathrm{c}}$ = 93 μA/μm². We increased the number of Josephson junctions as compared to the studies in Sec. V in order to increase the output voltage and, consequently, the signal-to-noise ratio of the EOS measurements.

In order to sample the output of an electrical device using our EOS platform, we have two possibilities: (i) We can synchronize the repetition rate of our probe laser to the repetition rate of the electrical device and use lock-in detection. In this case the output of the electrical device needs to be amplitude modulated at the lock-in detection frequency and time resolution is obtained by varying the time delay between the probe laser pulses and the electrical signal. (ii) We use an asynchronous sampling technique in which the repetition rate of the probe laser and the repetition rate of the electrical signal are either detuned from each other or harmonically related (but fixed otherwise). The signal can then be reconstructed by detecting each single laser pulse, if the repetition-rate difference between the probe laser and the electrical signal is exactly known [26].

For the measurements presented in this and the following section, we have used asynchronous sampling. To measure directly a JAWS waveform, we fixed the repetition rate of the probe laser to exactly 76 MHz and the frequency of the JAWS signal to exactly 1 MHz. We recorded 1672 million laser pulses over a measurement time of 22 seconds. We then adopted our data analysis such that we exactly record two periods of the JAWS signal with 2 x 76 = 152 data points in total. The corresponding result, displaying two periods of the 1 MHz JAWS output, is shown in Fig. 5(a). In a second measurement, we drove the JAWS with an electrical pulse pattern to produce a two-tone output with 1 MHz and 2 MHz frequency components and employed the same data analysis as before. The resulting two-period signal is shown in Fig. 5(b) and visualizes that also complex time-domain shapes can be sampled with our EOS platform. For both traces shown in Fig. 5 the y axes have been scaled to the expected JAWS output voltages.

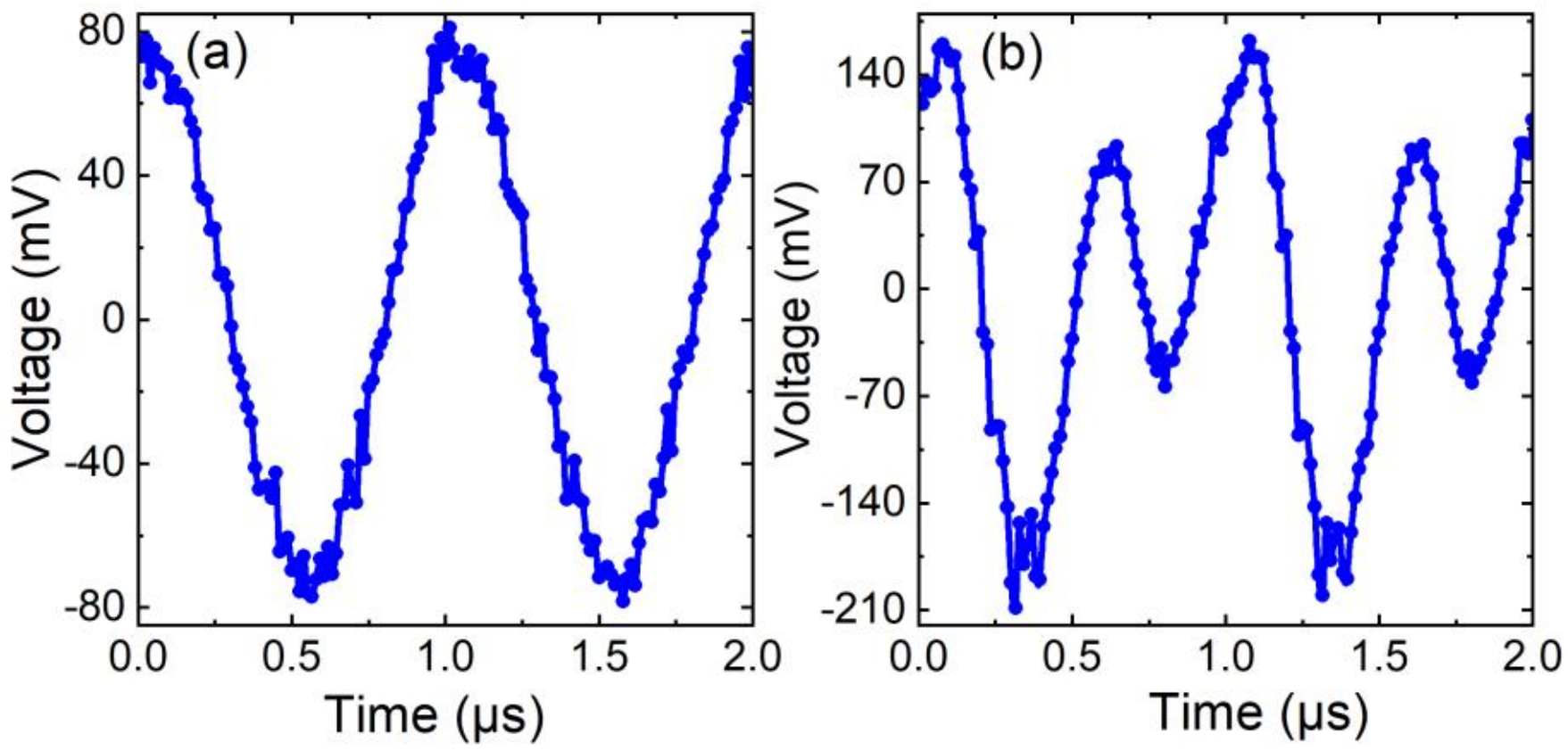

Fig. 5. EO measurement of JAWS (a) single frequency output at 1 MHz and (b) two-tone output with spectral lines at 1 MHz and 2 MHz.

The two traces shown in Fig. 5 look noisy. However, we like to emphasize that this is not white noise, as we cannot reduce it by simple averaging. Presently, we are investigating two aspects that might produce such features. (i) The bandwidth of the balanced detector used for the asynchronous measurements was 75 MHz. This also means that subsequent laser pulses cannot be fully separated from each other and might influence each other. (ii) The resolution of the ADC used in the asynchronous measurements was limited to 12 bits and this might lead to digitization errors. We are confident that further studies will improve the signal quality of our asynchronous EOS measurements.

## VII. Direct EOS of an optical pulse pattern generator

### *VII.1. Optical pulse pattern generator*

Josephson arbitrary waveform generators have recently been deployed to generate signals in the GHz regime by using electrical pulse pattern generators as drivers [27]. These instruments are limited to pulses of tens of picoseconds of width. Fast pulsed lasers feeding novel optical-to-electrical converters [28] with > 100 GHz bandwidth to drive Josephson junction arrays with similarly high characteristic frequency would allow for the generation of quantized pulse patterns in an unprecedented frequency range. To test such assemblies we have built an optical femtosecond pulse generator (OPPG) that can be used to produce optical pulse clusters with freely adjustable pulse intervals and amplitudes, see Fig. 6.

The core of the OPPG is a homemade, passively mode-locked 1530-nm Er-fiber laser with a pulse repetition rate that is adjusted to be an integer multiple of the EOS frequency: 15 x 76 MHz = 1.14 GHz, see Fig. 6(a). The average output power of the oscillator is approximately 1 mW. The pulses are amplified to about 200 mW in an erbium-doped fiber amplifier [29] and temporally compressed to about 230 fs. Finally, the spectrum is broadened in a highly nonlinear fiber and filtered by a 12-nm bandpass filter at 1480 nm, yielding about 5 mW average power. The filtered wavelength corresponds to the maximum responsivity of InGaAs p-i-n photodiodes at 4 K. The pulse duration after a couple of meters of additional SM fiber is around 420 fs (without dispersion compensation). The pulse repetition rate is phase-locked to the 15$^{th}$ harmonic of the EOS signal at 1.14 GHz by controlling the pump power. A piezo attached to the gain fiber is used to make slow adjustments to the cavity length in order to keep the average pump power at the nominal set point. Coarse repetition rate changes can be done by adjusting the cavity length mechanically or by adjusting the cavity temperature.

For increasing the pulse repetition rate, the pulses enter a polarization controller, a polarizer and are then fed into a polarization-maintaining fiber-optic frequency multiplier composed of 2x2 splitters and free-space delay lines leading to a pulse rate of 4.56 GHz, see Fig. 6(b). This regular pulse train is further coupled into four parallel free space delay lines, which enable adjusting the successive four pulses with any interval down to the limit when they start to overlap. Finally, we combine the interleaved pulses into a single fiber with a 4x1 splitter. In the following measurements, the optical pulses were let to broaden in a 5-m long SM fiber to around 1 ps (FWHM) before reaching the cryogenic PD.

### *VII.2. EOS measurements*

In Fig. 7 we show data measured by exciting the 60 GHz PD with different optical pulse trains at higher frequencies. Figure 7(a) and (b) contain EOS data with pulse-train frequencies of 4.56 GHz and 2.28 GHz, respectively. The optical pulses were first set to be approximately equal in power using an optical power meter. In the EOS data the main pulses resulting from optical excitation are clearly visible, marked with arrows. Since the secondary pulses originating from reflections at the impedance mismatches live longer than 1 ns, and are not synchronous with the excitations, the background level at each primary pulse is varying. It is also clearly seen that the main pulses are followed by different secondary pulse patterns.

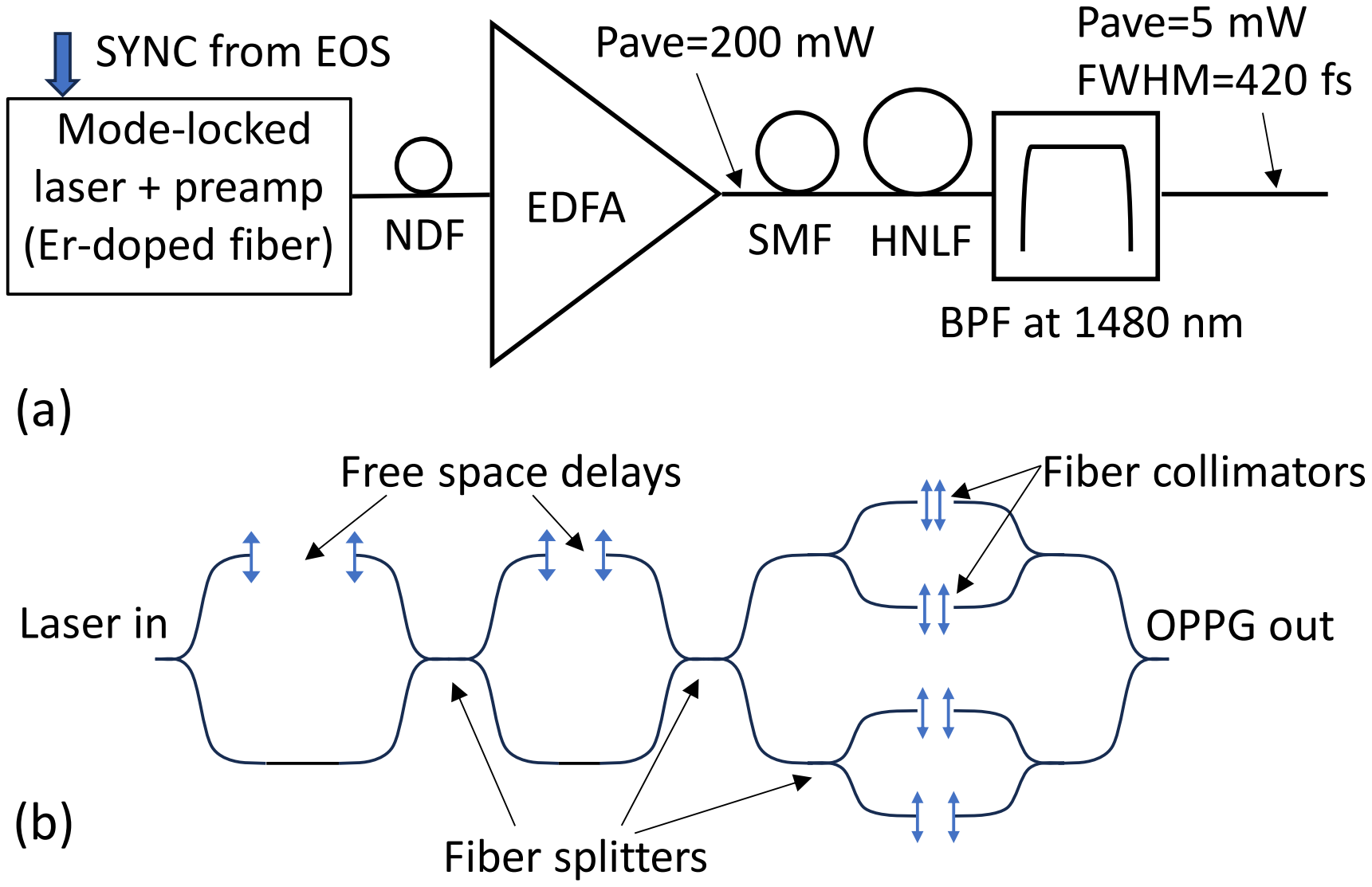

Fig. 6. (a) A custom-made femtosecond-pulse laser composed of a mode-locked fiber laser, optical amplifiers and spectral broadening section. $P_{ave}$ = 5 mW of average power at 1480 nm (in the 12 nm band) is available at 1.14 GHz repetition rate. EDFA denotes an Erbium-doped fiber amplifier, NDF a negative-dispersion fiber, HNLF highly nonlinear fiber, and BPF a band-pass filter. (b) Optical pulse multiplexer based on fiber splitters, fiber collimators and free-space delay paths.

In Fig. 7(c) we show four-pulse clusters with 55 ps interval between neighboring pulses and repetition period of 1/1.14 GHz. The optical pulses were again set equal in power. The signal background consisting of back and forth reflected pulses is even more pronounced than in the above cases. Looking at the main pulses closely we observe that the FWHM is only 3.6 ps, and there is no indication of a tail as observed with 200-fs-pulse excitation, see Fig. 2(a). In the inset of Fig. 7(c) we compare the second pulse of the trace shown in the main plot of Fig. 7(c) to a pulse generated with the 200-fs excitation of a similar PD in a Au CPW leading to 5.4 ps FWHM, see also discussion in Sec. III. We think that the key difference in the measurements is the exciting optical pulse length, 200 fs versus 1 ps, i.e., that increasing the temporal width of the optical pulse, and by this reducing the optical peak intensity, decreases the temporal width of the resulting electrical pulse. This supports our preliminary interpretation of electrical pulse broadening in case of strong fs-pulse excitation, see discussion in Sec. III.

The traces shown in Fig. 7 contain significant reflections. The two most prominent impedance mismatches are (i) the ends of the CPW and (ii) the $LiTaO_3$ crystal. As for (i) we like to point to Sec. IV and Fig. 3, where we have shown that bond pads and wires indeed lead to severe reflections in our setup. We even used the EOS to exactly measure the reflection coefficient of one end of the CPW. In order to avoid reflections from bond pads and wires, it might be necessary to integrate the optoelectronic pulse generators, i.e., photodiodes, with the corresponding device under test in one chip (e.g., flip chip bonding or monolithic integration) and use on-chip terminations. As for (ii) we like to point to two previous publications in which we investigated the distortions from a $LiTaO_3$ crystal versus its height above a coplanar transmission line [30] and experimentally determined a transfer function of a $LiTaO_3$ crystal allowing for the correction of EOS measurements [31]. Yet, the studies of Refs. [30,31] were done at room temperature and corresponding experiments in a cryogenic environment are much more difficult. In any case we believe that EOS is the most straightforward way to study reflections resulting from impedance mismatches in cryogenic circuits. In this regard, cryogenic EOS with the possibility of low-invasive in-situ measurements in superconducting circuits even outperforms room-temperature electronics working in the time domain (oscilloscopes) and frequency domain (vector network analyzers).

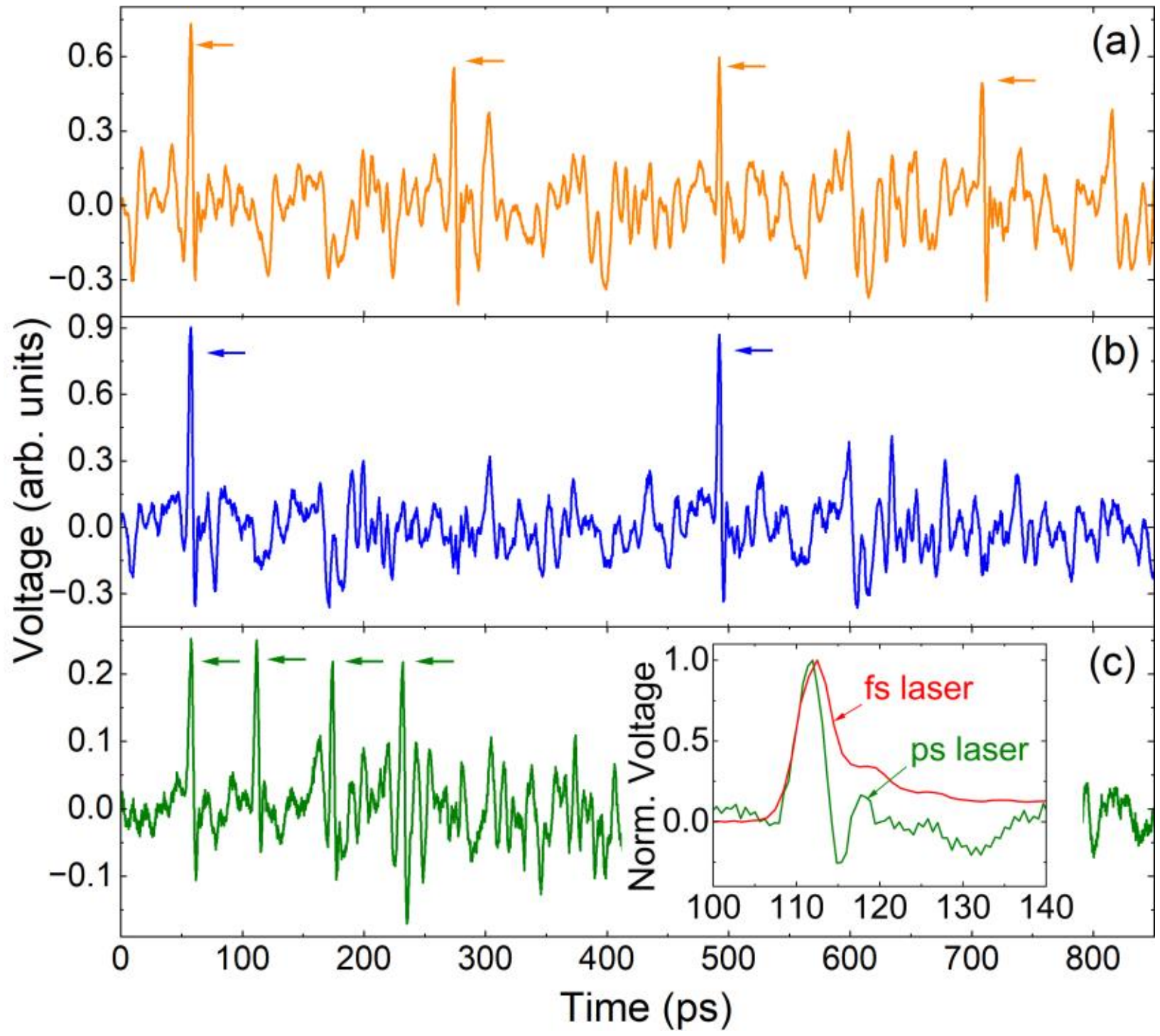

Fig. 7. EOS data measured with different optical pulse clusters (optical pulse FWHM=1 ps) used to drive a 60 GHz PD flip-chip bonded on a Nb CPW. The EOS measurement was carried out in the middle of a 13-mm long CPW between the PD and the wire-bond pads, which couple to a coaxial line in a cryoprobe. The horizontal arrows point to the temporal positions of the voltage signals generated by the optical pulse cluster. (a) 4 pulses equally spaced within a $(1.14\ \mathrm{GHz})^{-1}$ time epoch (~220 ps time interval between pulses). (b) 2 pulses equally spaced within a $(1.14\ \mathrm{GHz})^{-1}$ time epoch (~440 ps time interval between pulses). (c) 4 pulses equally spaced within a $(4 \times 1.14\ \mathrm{GHz})^{-1}$ time epoch (~55 ps time interval between pulses). The noise (standard deviation) of the experimental technique is approximately 0.016 in relation to the arbitrary units of the y axis. The inset compares primary PD pulses when excited with 1-ps optical pulses and 200-fs optical pulses.

The data presented in this section illustrates that our cryogenic EOS technique is well capable or resolving waveforms of complex and fast excitations. Combined with the optical pulse pattern generator principles presented above, this yields a tool to investigate performance of cryogenic ultrafast systems even up to frequencies exceeding 100 GHz.

## VIII. Conclusions

We demonstrated EOS as a method for time-domain measurements of ultrafast electrical signals in cryogenic and superconducting circuits. Our results show the importance of various impedance mismatches when aiming at delivering ultrafast electric pulses along coplanar transmission lines. Cryogenic EOS is capable of measuring high-frequency signals >100 GHz in situ in superconducting circuits in a cryostat, which is not possible with equal precision by any other means. The measurement technique can resolve voltages with better than 1 mV uncertainty [32]. We have also developed EOS based on an external electro-optic sampling tip attached to a cryogenic piezo stage, which enables setting the tip at different locations and thereby identifying forward and backward travelling waveforms in transmission lines. Our work also denotes a significant step for traceability in microwave measurements in superconducting circuits. We pioneer the direct calibration of electro-optic sampling against a quantum voltage standard, but more work is needed to deliver a detailed uncertainty analysis and to compare the calibration using a quantized voltage source to previous calibration methods [21].

We have also demonstrated a fast, optical pulse signal source suitable for driving optical-to-electrical converters, and further circuits such as JAWS or JPG. The optical source allows arbitrary setting of successive pulse amplitudes and intervals allowing various performance tests of JJ devices in the future. As an example of our measurements we were able to verify a commercial 60 GHz InGaAs p-i-n photodiode producing electrical pulses with only 3.6 ps FWHM at 4.2 K when excited with 1 ps optical pulses as compared to 5.4 ps FWHM for 200 fs long excitation pulses. For completeness we note that the excitation wavelengths are also different between the fs and ps excitation. To investigate these issues in more detail, we plan to perform a consistent study in which we broaden the femtosecond excitation pulses from ~100 fs to several ps while maintaining the same spectral bandwidth. This will reduce the peak power and, consequently, also reduce electric-field screening and absorption saturation in the depletion region. Such an experiment is currently under study but beyond the scope of the present paper.

In general, the techniques presented in this paper may help to characterize, optimize, and verify various cryogenic microwave components and circuits, such as photodiodes and various interfaces including flip-chip and wire bonds. Furthermore, optical methods such as optical pulse pattern generators and photodiodes may enable increasing driving JAWS or JPG with pulse frequencies approaching 100 GHz, but such devices require careful microwave optimization for which our EOS provides an unparalleled instrument. We expect that EOS sampling will become an elemental tool in studying the propagation of ultrafast signals in superconducting circuits, thus paving the way for new frequency bands in quantum technologies. Examples of research fields that might benefit from our technology are fast single flux quantum circuits like JAWS and novel high-frequency qubits [33].

**Acknowledgement**

This work was partly supported by the EMPIR programme co-financed by the Participating States and by the European Union's Horizon 2020 research and innovation programme (grant agreements 20FUN07 SuperQuant and 23FUN08 MetSuperQ), by the European Union's Horizon 2020 research and innovation programme (grant agreement 899558 aCryComm), and by the German Federal Ministry of Education and Research (grant agreement 13N15934 QuMIC). In addition, funding has been received from Research Council of Finland (grants 35220/QuantLearn and 359284/Finnish Quantum Flagship) and from an internal program of VTT.